\documentclass[twocolumn,showpacs]{revtex4}
\usepackage[dvips]{graphicx}
\usepackage{dcolumn}
\usepackage{amsmath}
\makeatletter
\def\btt#1{\texttt{\@backslashchar#1}}%
\DeclareRobustCommand\bblash{\btt{\@backslashchar}}%
\makeatother

\newcommand{\bra}{\left\langle}
\newcommand{\ket}{\right\rangle}

\begin{document}
\title{Dislocation nucleation in shocked fcc solids: effects of temperature and preexisting voids}
\author{Takahiro Hatano}
\affiliation{
Institute for Materials Research, Tohoku University, Sendai, 980-8577 Japan}
\date{\today}

\begin{abstract}
Quantitative behaviors of shock-induced dislocation nucleation are investigated 
by means of molecular dynamics simulations on fcc Lennard-Jones solids: a model Argon.
In perfect crystals, it is found that the Hugoniot elastic limit (HEL) is 
a linearly decreasing function of temperature: from near-zero to melting temperatures.
In a defective crystal with a void, dislocations are found to nucleate on the void surface.
Also HEL drastically decreases to $15$ percent of the perfect crystal 
when the void radius is $3.4$ nanometer.
The decrease of HEL becomes larger as the void radius increases, 
but HEL becomes insensitive to temperature.
\end{abstract}

\pacs{62.20.Fe,62.50.+p}
\maketitle

Mechanical properties of shock-loaded solids are important to materials science 
since they can reveal reliability of materials under extreme conditions.
Shocks in solids generate high pressure almost instantaneously, 
and create extremely high deformation (strain) rates.
Yielding phenomena realized at high strain rates are 
usually quite different from those of quasistatic ones.
Notably, Rohde has found that the HEL is almost independent of temperature \cite{rohde}; 
Kanel and his coworkers have found that the HELs of Al and Cu single crystals are 
increasing functions of temperature \cite{kanel}.
These experiments are in apparent contrast with quasistatic deformations, 
where yield strength considerably decreases with increasing temperature.
However, contrary to Rohde and Kanel, stainless steel shows that 
the HEL is a decreasing function of temperature \cite{gu}.
These puzzling results provide us with intriguing problems.
The microscopic foundations of plastic deformation are dominated 
by the dynamics of dislocations.
Unfortunately, the nature of dislocation dynamics is very complicated 
and yet to be fully understood from theoretical point of view.
Therefore, it is reasonable to decompose the problem into three essential ingredients 
of dislocation dynamics: nucleation, mobility, and multiplication.
As the first stage of this line of thought, 
the properties of dislocation nucleation are investigated in this paper.

In a perfect crystal, dislocations nucleate with the help of thermal fluctuations.
Holian and Lomdahl have studied the emergence of stacking faults initiated by 
partial dislocation emission in a shocked perfect crystal \cite{holian}.
Recently, Tanguy et al. \cite{tanguy} have performed extensive simulations 
to confirm that there exists a critical size of dislocation loop 
below which nucleated loops are energetically unstable and eventually annihilate; 
this is reminescent of droplet nucleation in a metastable gas or liquid.
Although these two studies are enlightening, the role of temperature, 
which is known to be important to dislocation nucleation, has not been considered.
In some studies, temperature plays no role in activating dislocation nucleation \cite{xu}.
However, dislocation nucleation seems to be anyhow a fluctuation-assisted process 
in the sense that no nucleation is observed in a molecular dynamics simulation 
of a perfect crystal at zero temperature \cite{holian}.
It is important to understand temperature effects on dislocation nucleation 
to clarify the controversial experimental results \cite{rohde,kanel,gu}.

This paper discusses temperature effects on dislocation nucleation 
through molecular dynamics simulations.
It turns out that the HEL is a linearly decreasing function of temperature from zero to melting temperatures.
Then a defect is introduced into the system as a preexisting void, 
at which dislocation nucleation is found to be enhanced.

The present simulations use a simple Lennard-Jones (LJ) potential 
$U(r)=4\epsilon\left[(\sigma/r)^{12}-(\sigma/r)^6\right]$ with a cutoff length of $4.0\sigma$.
To avoid a discontinutity at the cutoff, the potential is suitably shifted.
Note that simulations with shorter cutoffs such as $2.6\sigma$ resulted 
in a visible decrease of HEL, as much as $90$ percent.
Hereafter, we let $\sigma = 3.41$ \AA and $\epsilon=1.65\times10^{21}$ [J] 
in order to make a quantitative comparison with experiments on solid Argon.
The Lennard-Jones potential has been well-tested in terms of mechanical and thermal properties of Argon 
such as the elastic constants and the melting temperature.
Hence it is reasonable to adopt the LJ potential as the first computational attempt 
to understand the anomalous HEL behavior of fcc metals \cite{kanel}.

In our system the axes of $x$, $y$, and $z$ are taken to coincide with 
$\bra 100\ket$, $\bra 010\ket$, and $\bra 001\ket$ directions, respectively.
Shocks propagate along the $[001]$ direction.
As usual, periodic boundary conditions are applied to $x$ and $y$ directions.
The whole system consists of $100\times 100\times 200$ unit cells.
Shock waves in the present simulations are generated by pushing 
a piston of infinite mass into the still target. 
The velocity of the piston is denoted by $u_p$.
We call the piston velocity $u_p$ throughout this paper.

In this work, the HEL is defined as the longitudinal stress ($\sigma_{zz}$) 
above which dislocation nucleation is observed.
Note that this definition of HEL differs from the one that is based entirely on shockwave behavior: 
the shock front decomposes into an elastic precursor and a plastic wave.
However, these two definitions are identical 
because the decomposition is also observed in molecular dynamics simulations 
once dislocations are emitted in shocked solids.
Longitudinal stress is calculated from the Hugoniot relation;
$\sigma_{zz}= \rho_0 u_s u_p$, where $\rho_0$ and $u_s$ denote 
initial density and shock velocity, respectively.
Note that the initial pressure is zero.
The longitudinal stress $\sigma_{zz}$ is approximately proportional to 
the shear stress \cite{holian} which is directly responsible for dislocation emission.

\begin{figure}
\includegraphics[scale=1.1]{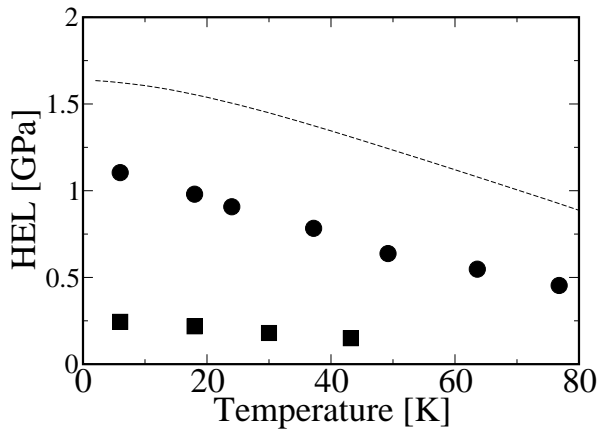}
\caption{Temperature dependence of Hugoniot elastic limits of a perfect crystal (circles) 
and of a defective crystal containing a void of $1.71$ nm radius (squares).
The dashed line represents experimental data for the shear modulus \cite{keeler}.
The melting temperature is approximately $83$ K.}
\label{hel}
\end{figure}
Figure. \ref{hel} shows the nearly linear dependence of HEL on temperature.
The HEL at $77$ K, close to the melting temperature ($83$ K) \cite{agrawal}, 
is approximately half that $6$ K.
This is clear evidence that dislocation nucleation is a thermally assisted phenomenon.
Also, this temperature dependence is reminiscent of elastic constants.
The dashed line in FIG. \ref{hel} represents the shear modulus of solid Argon \cite{keeler}.
We can see that the temperature dependences of the HEL and of the shear modulus 
are qualitatively the same for perfect crystals.
That is, HEL is approximately half the shear modulus regardless of temperature.
(Indeed, the ratio of HEL to shear modulus is slightly greater than $0.5$ for $T\le24$ K.)
The critical strain also decreases linearly from $0.138$ (at $6$ K) 
down to $0.118$ (at $77$ K) as the temperature increases.
Numerical data are shown in Table \ref{helvalues}.
Note that the temperature dependence of the shear modulus of Argon 
has been calculated by molecular dynamics simulation using LJ potential \cite{shimizu}.

\begin{table}
\caption{Critical piston velocity, Hugoniot elastic limit, and critical strain for the perfect crystal 
as functions of initial temperature.}
\begin{tabular}{c|ccc}
\hline
Initial & critical & HEL & critical \\
temperature & piston velocity &  & strain\\
K & m/sec & GPa & \%\\
\hline
6  & 291 & 1.08 & 13.8\\
18 & 275 & 0.957 & 13.7\\
24 & 261 & 0.886 & 13.1\\
37 & 243 & 0.765 & 13.0 \\
49 & 216 & 0.623 & 12.3\\
64 & 202 & 0.535 & 12.2\\
77 & 186 & 0.443 & 11.8\\
\hline
\end{tabular}
\label{helvalues}
\end{table}

The temperature dependence of HEL obtained by the present simulation 
is opposite to the experiment on fcc metals \cite{kanel}.
But this is not a contradiction, for the experiment involves real crystals 
where various defects and impurities preexist.
Plastic deformations of real materials are mainly governed by 
the motion of preexisting dislocations, while in a perfect crystal 
only the dislocation nucleation is considered here.
Hence, the mobility of preexisting dislocations must be analyzed for 
further understanding of temperature dependence of shock-induced plasticity.
This subject will be discussed elsewhere.

\begin{figure}
\includegraphics[scale=.4]{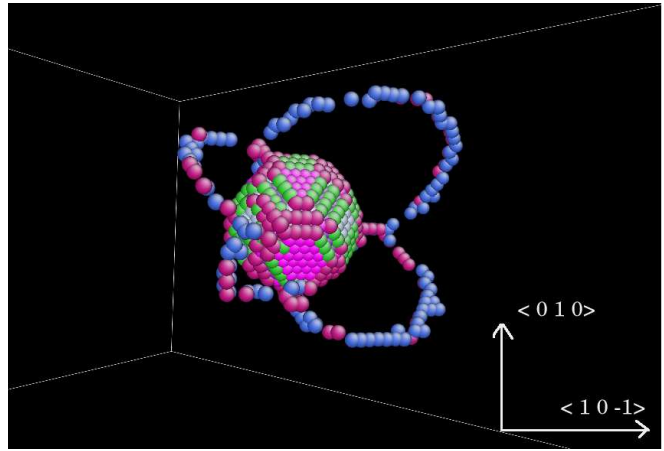}
\caption{Four partial dislocation loops nucleate on the void surface.
Void radius is $2.4$ nm and piston velocity is $70$ m/s.
Atoms are colored according to the number of the nearest neighbors; 
green, red, and blue atoms have $10$, $11$, and $13$, respectively.
Atoms of $12$ nearest neighbors are not presented \cite{li}.}
\label{nucleation}
\end{figure}

\begin{figure}
\includegraphics[scale=0.4]{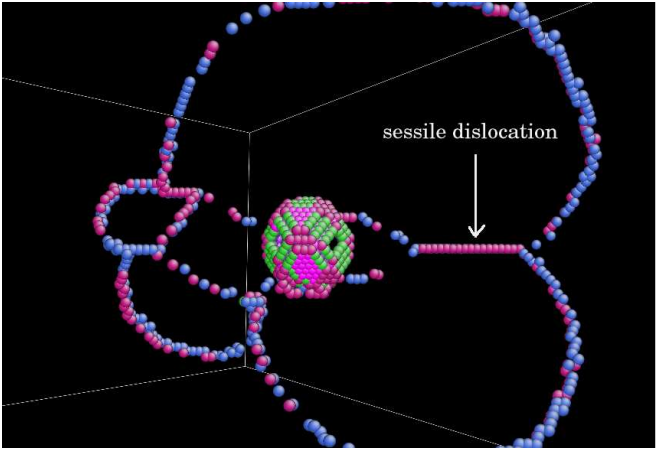}
\caption{Thirteen picoseconds after FIG. \ref{nucleation}. 
Partial dislocation loops are extendend and Lomer-Cottrel sessile dislocation is formed.}
\label{sessile}
\end{figure}

\begin{figure}
\includegraphics[scale=0.4]{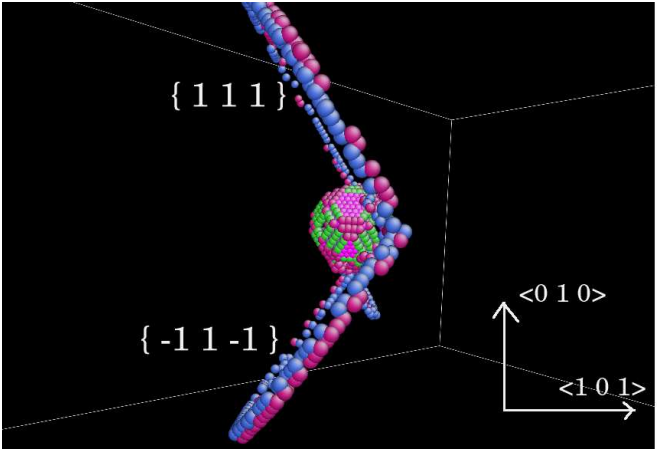}
\caption{Sideview of FIG. \ref{sessile}. Two slip planes on the void surface are activated.}
\label{sideview}
\end{figure}

Strikingly different results are found when we turn to defective crystals, 
concentrating on the effects of a void.
A void is introduced in a model crystal as a blanked spherical region.
The radius is varied from $0.34$ nm to $3.41$ nm.
With a void, dislocations nucleate exclusively on the void surface 
as shown in FIGS. \ref{nucleation}, \ref{sessile}, and \ref{sideview}.


The Hugoniot elastic limit decreases drastically in the presence of a void.
Square symbols in FIG. \ref{hel} represent the HEL for a defective crystal with a void of $1.71$ nm radius.
With this size of void, the HEL is approximately $20$ percent of the perfect crystal's.
Note that the HEL of a defective crystal also linearly decreases with increasing temperature up to $42$ K.

When $T>42$ K, approximately half the melting temperature, 
no dislocation is nucleated and a void collapses after the passage of shock front.
Although this phenomenon is similar to "hotspot" formation \cite{hatano}, this is not the case 
since the piston velocity used here is not strong enough to cause a jet which is essential to hotspot.
The collapse of the void observed in the present situation rather postulates 
that voids are unstable to perturbations above half the melting temperature \cite{gittus}.

Below $42$ K, the HEL and the critical piston velocity for the defective crystal 
with a void become insensitive to temperature.
For the case of $r=1.02$ nm, the critical piston velocity is $103\pm 5$ m/s at $6$ K, 
dropping only to $95\pm 5$ m/s at $42$ K, while for the perfect crystal 
the corresponding values are $289\pm 1.6$ m/s and $213\pm 1.6$ m/s, respectively.
The absolute decreases are $8$ m/s for the defective crystal and $76$ m/s for the perfect crystal.
Then the relative changes are $8/103\simeq 0.08$ and $76/284\simeq 0.26$, respectively.
The difference between the perfect and the defective crystals is remarkable 
in terms of both absolute and relative decreases.
Thus defects dominate thermal fluctuations in the enhancement of dislocation nucleations.
In experiment, the microstructure of the specimen is important.

\begin{figure}
\includegraphics[scale=1.1]{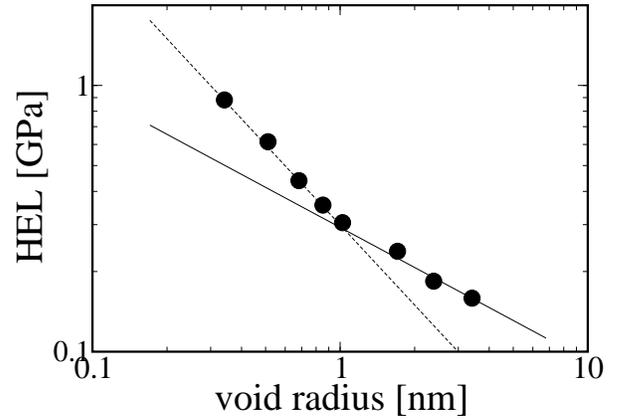}
\caption{Log-log plot of HEL as a function of void radius.
Solid and dashed lines are proportional to $r^{-0.5}$ and $r^{-1}$, respectively.
Initial temperature is $6$ K.}
\label{voidHEL}
\end{figure}

The void-size dependence of HEL is shown in Fig. \ref{voidHEL}.
At small $r$, the HEL decreases as $r^{-1}$ followed by a crossover to $r^{-0.5}$ at $r_c = 1.3\pm0.3$ nm.
Note that the transition radius $r_c$ is approximately the width of a double kink 
and is also very close to the critical loop radius $1.5$ nm 
above which nucleated dislocation loops grow stably \cite{tanguy}.
The quantitative explanation of this intriguing transition is not clear at this point.

\begin{table}
\caption{Comparison of simulation and experimental shock velocities 
as functions of piston velocity. (Simulation data involves a perfect crystal).
Initial temperature and density for the simulation are $64$ K and $1.63$ g/cc, respectively, 
while they are $75$ K and $1.65$ g/cc in the experiment.}
\begin{tabular}{c|cc}
\hline
Piston velocity & \multicolumn{2}{c}{Shock velocity} \\
      & (experiment) & (simulation) \\
km/sec & km/sec & km/sec \\
\hline
0.56 & 2.00 & 2.35 \\
0.78 & 2.44 & 2.77 \\
0.94 & 2.68 & 3.01 \\
1.29 & 3.53 & 3.82 \\
1.78 & 4.17 & 4.82 \\
\hline
\end{tabular}
\label{hugoniot}
\end{table}
To compare the present simulation with experiment, the Hugoniot values ($u_p$ versus $u_s$) of Argon 
for both this simulation and experiment \cite{dick} are presented in TABLE \ref{hugoniot}.
We see that the simulation values exceed the experimental ones by approximately $15$ percent.
This overestimation is comparable to that of Belonoshko's simulations using a Buckingham potential \cite{belonoshko}.
His deviation from the experimental value exceeds $10$ percent, 
especially for small piston velocities up to $1.2$ km/sec.
Dubrovinsky has attributed this overestimation to the polycrystalline nature of the experimental specimen \cite{dubrovinsky}.
Since both Belonoshko's and the present simulations deal only with crystals of $\bra 100\ket$ orientation, 
the deviation from the experimental result is not unreasonable.

Finally, two quantitative uncertainties in the present simulation are remarked.
One is underestimation of stacking fault energy by the use of LJ potential.
It is known that stacking fault energy vanishes when one adopts short-range, 
two-body intermolecular potentials \cite{dubrovinsky}.
In this regard, it is possible that the numerical values of the HEL 
obtained by the present simulations are underestimates.
But, at least, qualitative results such as the temperature dependence or the void effect 
are not so influenced by the nature of two-body potentials.
In addition, a large cut-off, $4.0\sigma$, was used to reduce this effect.

The other uncertainty invloves periodic boudary conditions.
The system size ($57\times57\times114$ nm) is small enough that 
a dislocation loop may interact with itself through the periodic boudary conditions. 
It is impossible to remove this uncertainty from molecular dynamics simulations at this point.
However, the strain around a nucleated dislocation loop propagates at the sound velocity 
(about $900$ m/s for transverse mode), taking $45$ picoseconds to cross the simulation cell.
Therefore, this efect may not seriously influence the present results, 
which involve exclusively nucleation processes that occur within $10$ picoseconds.

To summarize, molecular dynamics simulations on both perfect and defective fcc crystals 
are performed from near-zero to melting temperatures.
It is found that the Hugoniot elastic limit is about half the shear modulus 
and decreases linealy with increasing temperature.
The critical piston velocity and critical strain also decrease in the same manner.
When a void is introduced in the crystal, dislocations are found to nucleate on the void surface.
The critical piston velocity decreases drastically as the void radius $r$ increases.
The decrease is nonlinear and shows a crossover from $r^{-1}$ to $r^{-0.5}$ around $r \simeq 1.3$ nm, 
which is approximately the width of a double kink.
The Hugoniot elastic limit becomes insensitive to temperature in the presence of a void.
That is, the effect of nucleation at void dominates that of thermal fluctuations 
in shock-induced dislocation nucleation.

The author gratefully acknowledges helpful discussions with 
H. Kaburaki, F. Shimizu, Y. Satoh, N. Nita, and H. Matsui.

\end{document}